# Simultaneous Surface Plasmon Resonance and X-ray Absorption Spectroscopy

A. Serrano[1,2,*], O. Rodríguez de la Fuente[2], V. Collado[3], J. Rubio-Zuazo[3], C. Monton[5], G. Castro[3] and M. A. García[1,5]

[1] Instituto de Cerámica y Vidrio (ICV-CSIC), Cantoblanco, 28049 Madrid, Spain

[2] Departamento de Física de Materiales, Universidad Complutense de Madrid, 28040 Madrid, Spain

[3] SpLine, Spanish CRG Beamline at the ESRF, F-38043 Grenoble, Cedex 09, France and Instituto de Ciencia de Materiales de Madrid, (ICMM-CSIC), Cantoblanco, 28049 Madrid, Spain

[4] Department of Physics and Center for Advanced Nanoscience, University of California San Diego, La Jolla, CA 92093 USA

[5] IMDEA Nanociencia, Cantoblanco, 28049 Madrid, Spain

Corresponding author: aida.serrano@icv.csic.es

## Abstract

We present an experimental set-up for the simultaneous measurement of surface plasmon resonance (SPR) and X-ray absorption spectroscopy (XAS) on metallic thin films in a synchrotron beamline. The system allows measuring *in situ* and in real time the effect of X-ray irradiation on the SPR curves to explore the interaction of X-rays with matter. It is also possible to record XAS spectra while exciting SPR in order to study changes in the films induced by the excitation of surface plasmons. Combined experiments recording simultaneously SPR and XAS curves while scanning different parameters can be also carried out. The relative variations in the SPR and XAS spectra that can be detected with this set-up ranges from $10^{-3}$ to $10^{-5}$, depending on the particular experiment.

## 1. Introduction

Surface plasmon resonance (SPR) is an outstanding property of metallic nanostructures. It consists of collective oscillations of conduction electrons at metal/dielectric interfaces. In the case of thin films, SPR corresponds to transversal oscillations that propagate along the interface [1,2,3]. For noble metal films, SPR can be excited with visible light using the appropriated geometry. The excitation of SPR in these films concentrate and amplify locally the electromagnetic field of the light up to a factor of 80 [1,4]. Thus, SPR has many applications in different fields as sensoring [5], biomedicine [6] or optoelectronics [2]. Moreover, the features of SPR are extremely sensitive to modifications in the properties or morphology of the metallic film and the dielectric medium. Therefore, SPR spectroscopy is commonly used as probe to study in-situ the growth or modifications of dielectric materials [4]. Similarly, X-ray absorption spectroscopy (XAS) is an invaluable technique that provides detailed information on the electronic structure of materials [7,8]. It is also known that X-rays induce modifications in many kinds of materials. Therefore, the combination of both techniques may be used to study the interaction between electromagnetic radiation and matter by using one of the beams to modify the material and the other one as a probe. In spite of the very different energy ranges of SPR and XAS (~eV for SPR and ~keV for XAS), both processes deal with the excitation of electrons to about the Fermi level. Since SPR alters the electronic population of the metallic film at the Fermi level, a modification of the XAS spectra should be expected. Similarly, when electrons are pumped to the Fermi level by X-ray absorption, one might expect variations in the SPR spectra. Given that SPR is very sensitive to any modification of the dielectric medium close to the metallic film, it can be used also to study the permanent effects induced by X-rays on a large variety of materials as glasses, crystalline oxides or organic matter.

The combination of both techniques is technologically challenging. The dispersion relation of SPR and light do not cross each other. Thus, SPR can not be excited by direct illumination of the interface. The only chance to excite optically SPR it taking advantage of evanescent fields [1]. Evanescent fields allow an electromagnetic wave to penetrate and propagate few nanometres into a medium with the wavevector different to that determined by optical laws. Using these evanescent fields and for certain incidence angles, it is possible to match the dispersion relation of the light with that of the SPR at the interface. Therefore, optical excitation of SPR can be performed only with very special geometries rending very difficult its combination with other techniques. The fact that the sample rotates during the SPR measurements arise additional difficulties to align properly both the laser and X-ray beams. We have designed and tested a SPR system compatible with a XAS beamline. With this set-up we have explored the combination of both techniques and tried to determine the type of experiments that can be carried out and the information we can extract from them. The device has been mounted at the Spline BM25 beamline at European Synchrotron Radiation Facility (ESRF) in Grenoble, France, and is now available for experiments. We describe here the set-up and its capabilities.

# 2. Experimental

## *2.1 Experimental set-up*

The set-up was mounted on the branch A of the CRG BM25-SpLine Beamline at the ESRF [9,10]. This beamline is split into two branches A and B, each of them fully equipped with focusing optics and experimental stations that are operated independently. Branch A enables the performance of X-Ray Absorption Spectroscopy and High Resolution Powder Diffraction measurements. This branch is located on the soft edge of the D25 bending magnet with a critical energy of 9.7 keV and energy resolution of $\Delta E/E=1.5\text{x}10^{-4}$. The X-ray energy ranges between 5 and 45 keV and the flux is of the order of $10^{12}$ photons/s at 200 mA ring current. The beam spot size can be changed in all the energy range between 300x100 $\mu m^2$ and 40x10 $mm^2$ in the horizontal and vertical directions, respectively. The position and dimensions of the focused beam are kept constant during a ~ 1 keV energy scan, which represents standard conditions for Extended X-ray Absorption Fine Structure (EXAFS) measurement. The XAS system is arranged onto an optical table for optimum placement and alignment of the environmental sample and the detection equipment components. The system is equipped with motors covering all degrees of freedom, three translation stages (X, Y, Z) and three rotation stages for centring the sample. The precision of such rocking cradle motors is 0.001º and can tilt the sample stage within a range of ±15 º along the directions parallel and perpendicular to the incident beam. While the beamline can be operated in both transmission and fluorescence modes, only this later is used in our new set-up. A gas ionization chamber working in the low-pressure range (from OKEN) is placed for monitoring the incoming beam intensity, $I_0$. Fluorescence is detected with a nitrogen cooled 13-element Si(Li) detector from e2v Scientific Instruments. The low temperature of the Si(Li) and FET ensemble also diminishes leakage current and electronic noise. The Si(Li) fluorescence detector allows to perform fluorescence measurements in a range between 3 and 30 keV. The energy resolution at 6 μs peaking time is around 140 eV in every crystal. The detector is also fully equipped with translation and rotation motors to optimize its position, reducing in this way the solid angle for elastic diffuse scattering contributions.

Figure 1a show a scheme of the SPR set-up. Figures 1b and 1c displays pictures of the system mounted on the beamline. There are two possible configurations for the optical excitation of SPR, the Raether-Kretschmann [1,11,12 ] and the Otto ones [1,13] . This later requires placing a transparent medium with refraction index higher than 1 (typically a glass) tens of nanometres over the surface of the metallic film. It is therefore impossible to illuminate with X-rays the same region where the SPR is excited. Thus, we must use the Raether-Kretschmann configuration for which the metallic film grown over a transparent substrate is coupled to a prism on the glass side (see inset in figure 1a). The glass/metal interface is illuminated with a laser beam in total reflection conditions. An evanescent field propagates through the metallic layer reaching the other side where surface plasmons are excited. This configuration limits very much the thickness of the metallic film to a narrow window around ~50 ±10 nm [1,4]. For thicker films the evanescent film reaching the other side of the metallic film is very weak (it decays exponentially with the distance to the glass/metal interface). On the other hand, for films thinner than 50 nm the electron oscillations are damped.

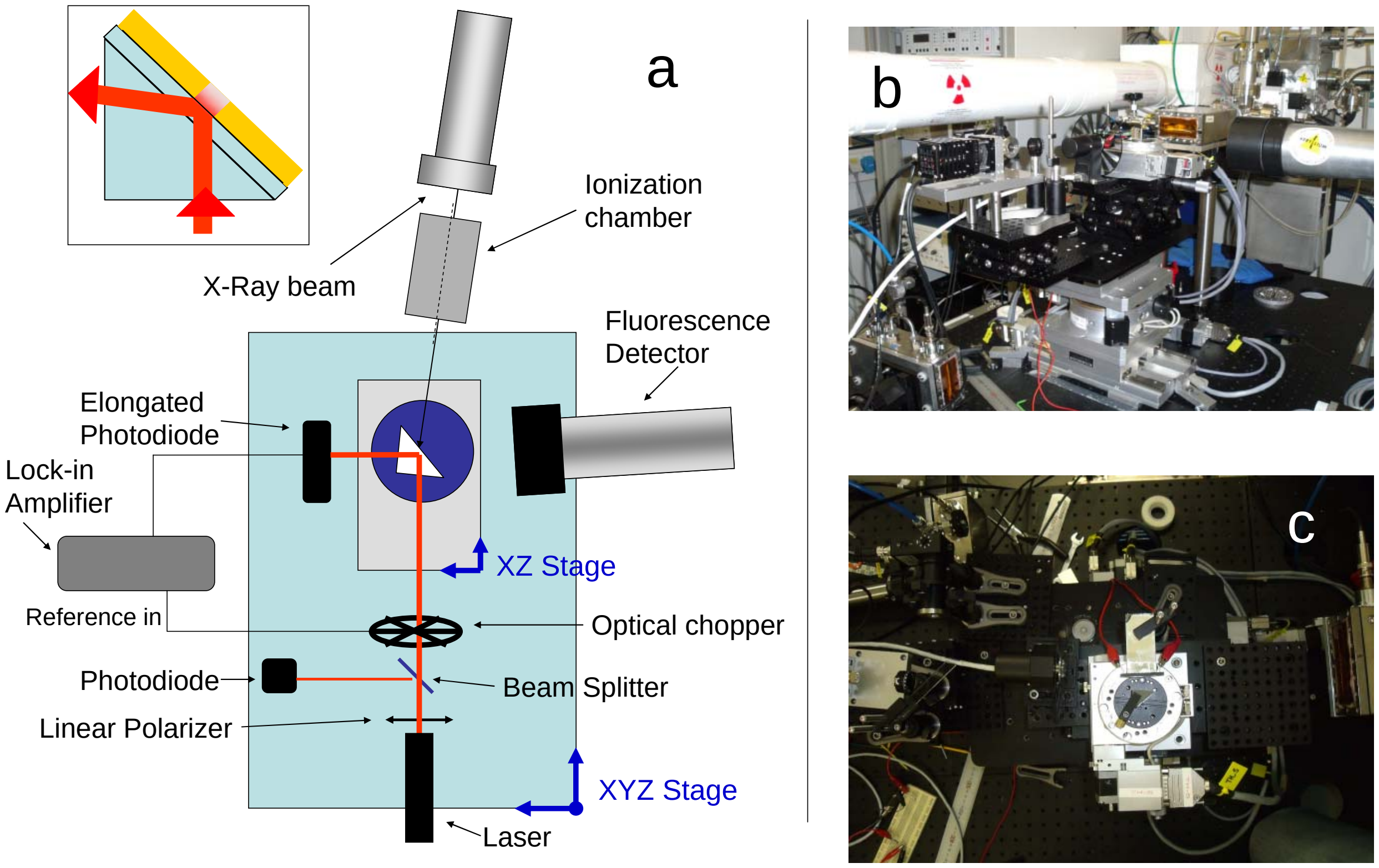

Figure 1. (Color online) (a) Scheme of the device for SPR and XAS measurements. Inset shows a detail of the glass/metal interface illumination. (b) Lateral and (c) top view photographs of the SPR device mounted in the experimental hutch of the SpLine beamline.

In our device, the SPR system is mounted onto the optical table of the XAS system. The translation and rotation motors enable positioning the whole SPR device system precisely with respect to the X-ray beam. Excitation is performed with a HeNe (632.8 nm) linearly polarised laser (power stabilized, CW 0.5 mW); other lasers are available. The laser is mounted on a cradle equipped with yaw and pitch movements for a precise orientation of the beam on the sample. A beam splitter deflects about 5% of the laser intensity to a photodetector in order to record fluctuations in laser intensity during the experiments. These fluctuations along several hours resulted below 0.05%. Beyond the beam splitter, the laser beam is modulated with an optical chopper working at 479 Hz. The sample, consisting of a thin metallic film (typically 50nm Au or Ag) grown on a glass substrate plus possible overlayers of dielectric materials, is fixed to a triangular (or semicircular) glass prism using gel index matching for a good coupling. Sample and prism are mounted on top of a rotating motor that allows changing the laser incidence angle. This sample stage has an independent XZ translation stage for sample positioning. During the measurements, the laser beam reflected at the sample is collected as a function of the incidence angle with an elongated photodiode (to avoid moving the detector during the angle scan). The photodiode signal is registered with a lock-in amplifier using the optical chopper as reference. All the elements that comprise the SPR device (motors, laser monitor and lock-in amplifier signals) are integrated and synchronized in the software for the control of X-rays system. To avoid that the X-ray passing through the optical prism could reach the laser cavity, the laser and X-ray beams are slightly tilted (~10º) as shown in figure 1. Note that, for example, for 14 keV X-rays about 0.1% of the radiation is transmitted through 5mm of silica. This small fraction may have significant effects in the instrumentation due to the high intensity of the X-ray beam. Actually, we observed that without tilting both beams, the X-ray photons reaching the laser induced thermal fluctuations leading to variations in laser intensity of the order of 0.5% despite it is power stabilized.

### *2.2 Alignment of the beams*

SPR spectroscopy in the Kretschmann-Raether configuration is performed by collecting the reflected light as a function of the incident angle by rotating the sample (rocking curve) [1]. In this configuration, only the points along the rotation axis remain in the same position when the sample rotates (see figure 2). If the laser spot does not impinge at the sample on this rotation axis, its position will change during the scan.  This is not critical in standard SPR experiments with homogeneous samples. However, for our experiments, if the laser and X-ray beam are aligned out of the rotation axis they will become misaligned as the sample rotates, illuminating different points at the sample surface, as illustrated in figure 2. Thus, it is crucial to ensure that the laser spot illuminates the sample surface at the rotation axis. This can be confirmed in a simple way using two lasers. If they are properly aligned over the rotation axis, no misalignment will take place when the sample is rotated. To avoid misalignment, circular prisms are preferable since for triangular prisms the refraction at the prism surface will induce a certain horizontal deviation of the laser beam which is dependent on the incident angle. For 1cm side triangular silica (n=1.52) prism, when scanning the incidence angle in a range of 10 deg, the spot will move about ~0.6 mm.

Once the laser is properly aligned at the film surface, the X-ray beam must be aligned over the same position. This alignment is performed by moving the whole SPR system with respect to the X-ray beam using the XAS system translation stage. It is not possible to observe directly the X-ray beam (experimental hutch must be closed before opening the beam). Therefore, the alignment can be performed using a material sensitive to X-rays such as a photosensitive paper placed at the sample position and checking after irradiation if the laser spot beam matches the spot induced by the X-rays. Alternatively, the alignment can be carried out taking advantage of the fact that hard X-rays induce darkening of some glasses [14]. After irradiating for few minutes, the glass exhibits a dark spot that can be also used to check the proper alignment with the laser. Once the laser and X-ray beam are aligned, we can reach different regions of the sample using the sample Z translation stage of the sample. Laser and X-rays beam will remain aligned over a different point of the sample but, since the displacement is vertical, this new point will still be in the rotation axis. With this system it is also possible to change from one sample to another without modifying the alignment.

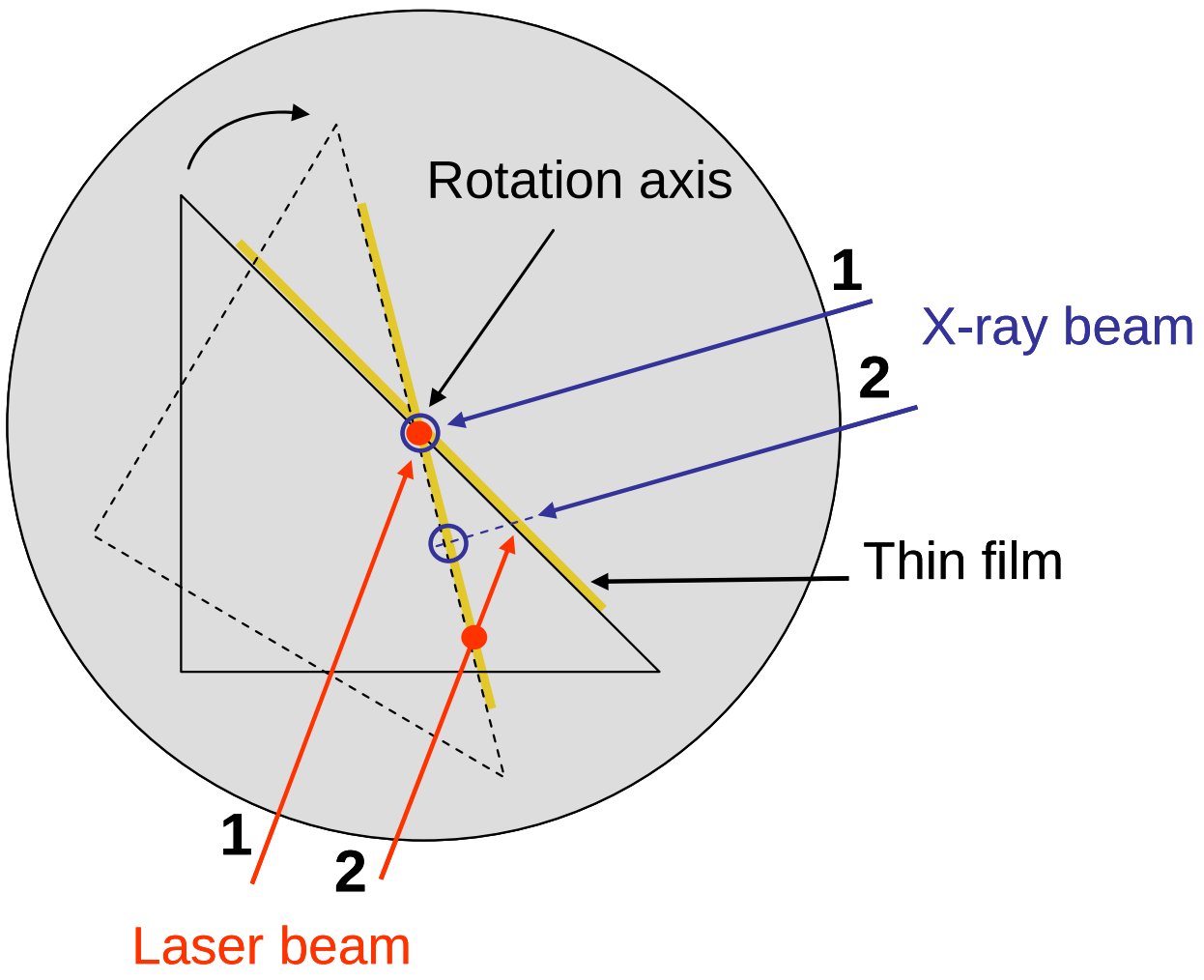

Figure 2 (Color online) Illustration of the possible misalignment of the laser and the X-ray beam. After the rotation of the prism, both beams illuminate the surface of the prism at the blue empty circle and at the red filled circle, respectively. If both beams are coincident at the rotation axis (case 1), they will remain aligned after the rotation of the prism (dashed prism). If the beams are initially not aligned over the rotation axis (case 2), they will intersect the prism surface at different points after the rotation of the prism.

To improve the quality of the measurements, it is important to ensure that the spot of the excitation beam is larger than the spot of the probe. To study the effect of X-ray irradiation on the SPR curve, the X-ray spot must be larger than that of the laser, so that the SPR spectrum is collected from a region fully illuminated with X-rays. In a similar way, to study the effect of SPR excitation on the XAS spectrum, the spot of the X-ray beam must be smaller than the laser spot, to ensure the recorded XAS spectra comes from a region where SPR is excited.

## 3. Results

We present now experimental results obtained with this sep-up to illustrate its capabilities.

Figure 3a shows two consecutive SPR spectra measured on a 50nm Au film grown onto a glass substrate. The difference between these spectra collected in the same conditions without X-ray irradiation (figure 3b) results of the order of $10^{-3}$ for the region of the spectrum with highest slope and of the order of $10^{-4}$ for the rest of the spectrum. With this equipment we can clearly detect relative differences below 0.1% with single scans. This resolution can be improved upon scans accumulation (scans in figure 3 take half an hour each one). We neither found a drift in the resonance position for ten consecutive scans up to a resolution of 0.01º, which is the minimum shift we could detect with this set-up.

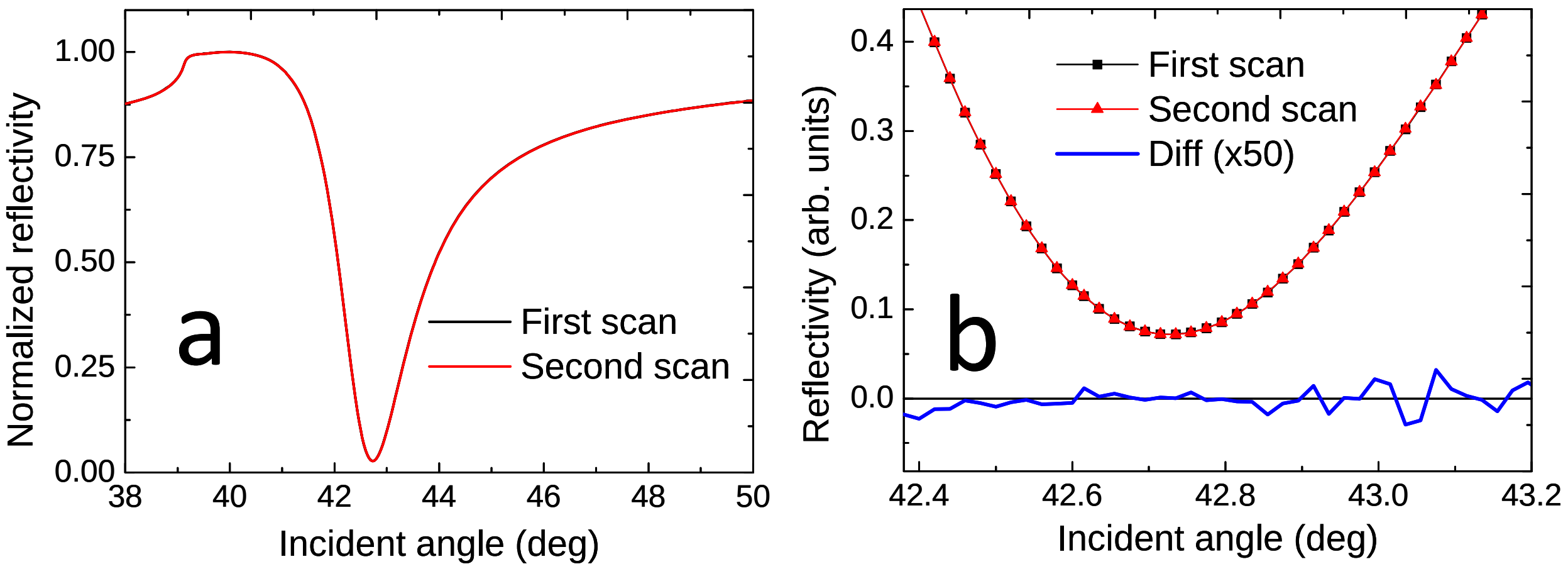

Figure 3 (Color online) (a) Consecutive spectra obtained with the SPR device shown in figure 1; the curves overlap and are not distinguished in the graph. (b) A detail of the resonance region and their difference (multiplied x50).

Figure 4 illustrates the effect of X-ray irradiation (E=11.952 keV, above the Au L3 edge) on the SPR spectra for Au films grown onto different glasses. For sodalime substrates (figure 4a), X-rays irradiation induces a decrease in the intensity of the reflected beam for the whole curve. The effect was found to be accumulative upon several scans. This effect is not related to SPR but to the darkening of the glass substrate upon X-ray irradiation. During the measurements, the laser beam travels across the glass substrate (see inset in figure 1a) and it is partially absorbed, reducing the intensity of the reflected beam. This darkening is observable with the naked eye after few minutes. It is well known that hard X-ray irradiation induces the appearance of colour centres in sodalime glasses [14,15,16]. These colour centres are commonly associated with network formers or modifiers (as Na or K in sodalime and B in the borosilicates) that change their coordination and oxidising state upon irradiation, leading to the formation of colour centres responsible of the glass darkening [17]. Besides the darkening, the X-rays induce a shift toward lower angles (0.02º) of the resonance as the inset in figure 4a shows. This modification corresponds to a decrease of the real refractive index of the glass [18,19]. A numerical analysis of the curve and comparison with simulations confirmed this shift corresponds to an increase of the refraction index of 0.001, that can be clearly resolved with the set-up.

With this set-up we can study the kinetics of the colour centres formation and elimination in real time. Fixing the incidence angle at a certain value and recording the reflectivity as a function of time while switching on and off the X-rays we may follow changes in the concentration of colour centres as shown in figure 4b. The kinetics of this process does not follow a simple exponential law and is strongly dependent on the X-ray energy. Actually, for X-rays with energy in the range between 6 keV and 9 keV, the darkening is only observable after several hours of irradiation. Note that reduction of the reflected beam intensity shown in figure 4a has nothing to do with surface plasmons. It is just due to the fact that the laser beam propagates along a medium that absorbs the light. Actually, this measurement can be carried out without any metallic film, by measuring the light transmission across a glass slide when it is irradiated with X-rays. The only advantage of the presented set-up is that we can measure simultaneously the variations in the real part of the refractive index (that shifts the resonance angle).

Contrary to the case of sodalime, for silica substrates (figure 4c) we found no significant variations in the reflectivity upon X-ray irradiation. However, a detail of the resonance region (figure 4d) upon X-ray irradiation reveals again a small decrease of the reflectivity and a shift of 0.02º toward higher angles upon X-ray irradiation. The shift corresponds to a decrease of the real part of refractive index of 0.001 induced by the irradiation. Therefore, the system allows monitoring changes induced by X-ray irradiation in both the real and imaginary part of the refractive index and the kinetics of the process with great accuracy.

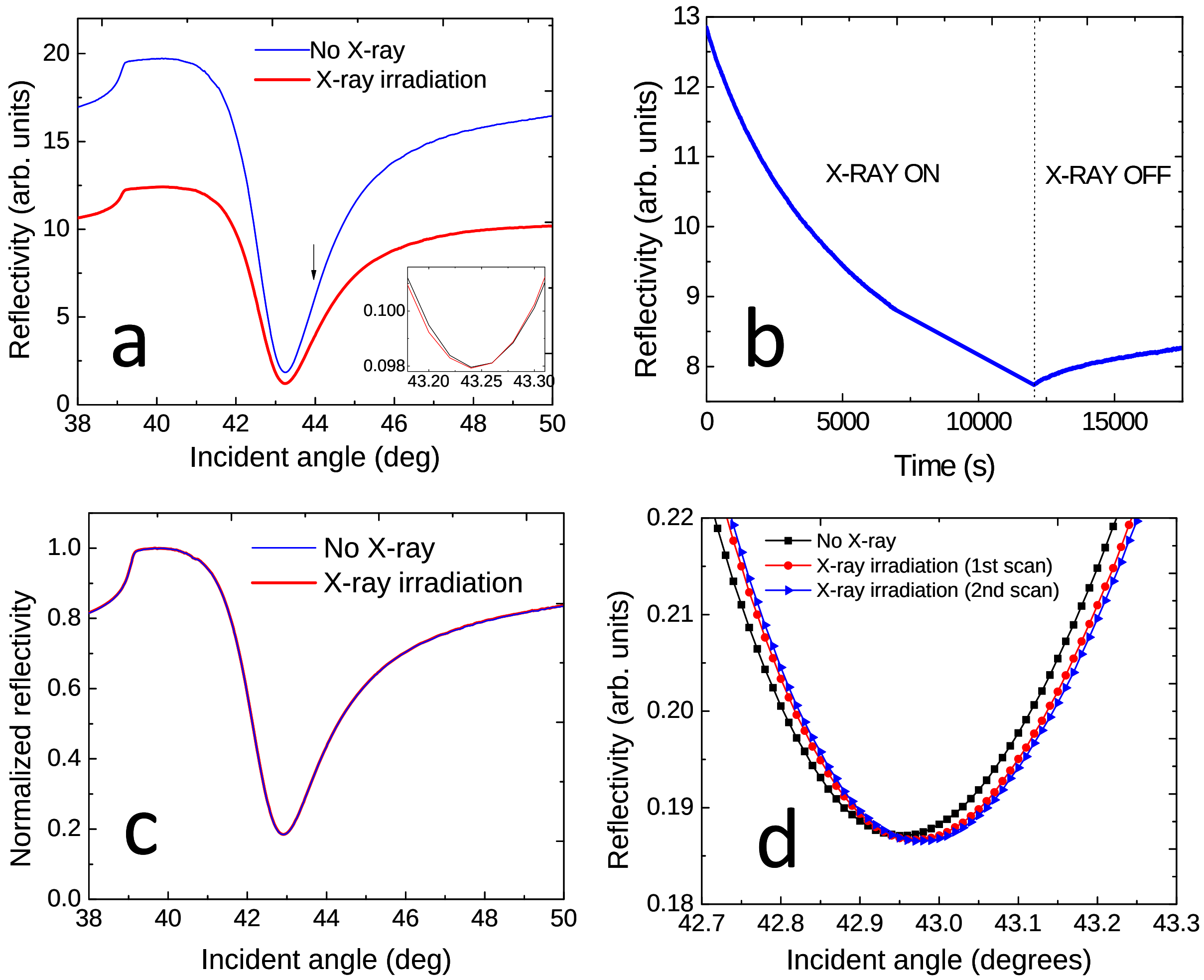

Figure 4. (Color online) (a) SPR curves for 50 nm Au film onto sodalime glass before and during irradiation with 11.952 keV X-rays. (b) Time evolution of the reflectivity at an incidence angle of 44º as a function of time during and after X-ray irradiation. (c) SPR curves for 50nm Au film onto silica before and during irradiation with 11.952 keV X-rays. (d) A detail of the resonance region.

As abovementioned, SPR is extremely sensitive to the features of the dielectric medium over the metallic film [1,4,5]. Thus, the setup can be also used to study the effect of X-ray irradiation on organic and inorganic dielectric layers grown on top of the metallic film. Figure 5 shows the SPR spectra for a 5 nm Co-Phthalocyanine (CoPc) layer over 50 nm Au film deposited onto a sodalime substrate [20]. Upon irradiation with X-rays at the Co K- edge (7.720 KeV) we observe a decrease of the intensity and a shift toward lower angles (about 0.2º) at the resonance, which is related to the damage induced by the X-rays on the CoPc film. The same experiment was repeated in a sample with identical glass and Au film but without CoPc film that did not show this phenomenology, confirming that it is due to modifications of the CoPc film.

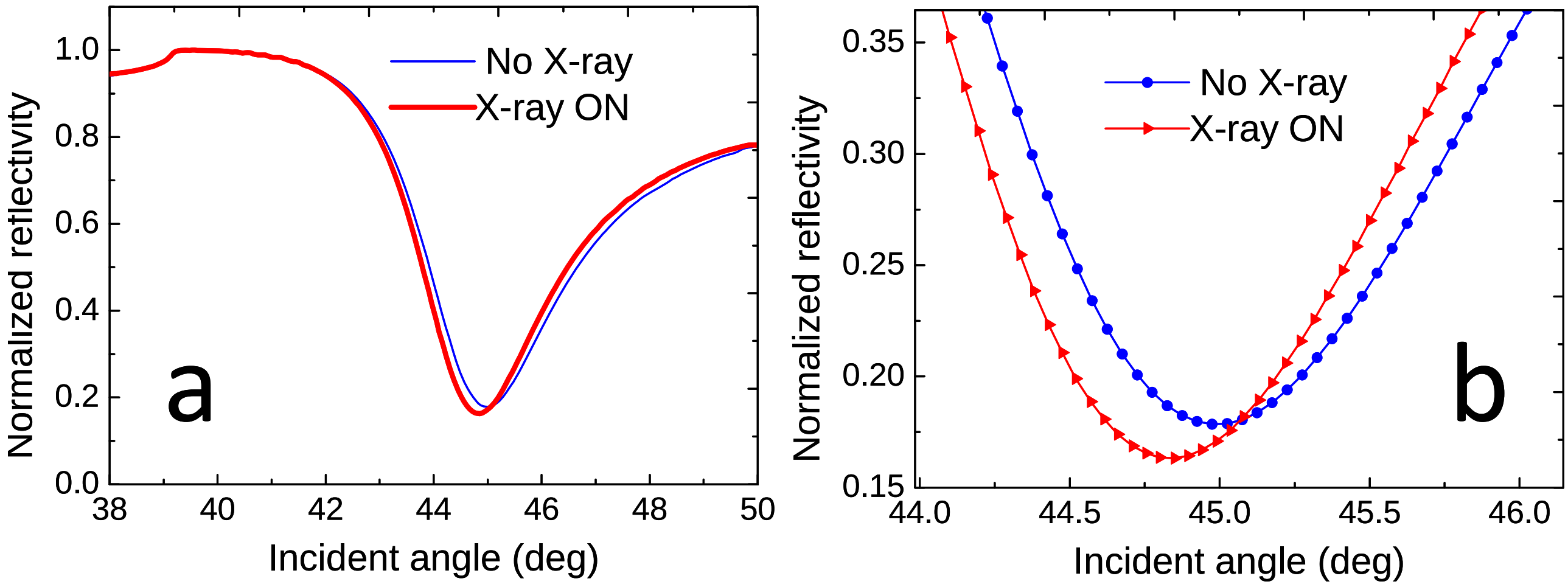

Figure 5. (Color online) (a) SPR spectrum of a 5nm Co-Phthalocyanine / 50 nm Au deposited on a sodalime substrate and (b) detail of the resonance region before and during irradiation with X-rays at the Co K edge (7.720 keV).

The examples above described correspond to modifications in the structure of the material induced by the X-ray which are accumulative and mostly permanent. The set-up allows also studying changes in the electronic structure (changes in the electron population at the Fermi level) due to X-ray absorption that are instantaneous and reversible. To achieve the best possible resolution and separate the instantaneous effects from the permanent ones, we have used the configuration illustrated in figure 6a. In this case the laser reaches the sample without any modulation while the X-rays are modulated with a chopper. The signal induced by the reflected laser beam at the photodiode is collected with the lock-in amplifier using as reference that of the chopper. Since the laser is not modulated, in absence of X-rays the signal measured at the lock-in is zero. If the X-rays induce any instantaneous and reversible modification of the material, the reflected beam will exhibit a component with the frequency of the chopper that will be detected by the lock-in. We have used this set-up to study possible electronic effects for an Au film grown onto silica when irradiating with X-rays at the Au L3 edge (11.914 keV). As shown in figure 6b there is no detectable effect up to a relative value of $10^{-5}$ of the SPR signal, which is the sensitivity of the set-up (the background value is independent of the X-ray irradiation). This result is consisting with calculations on the conduction electron density in Au at the Fermi level. Considering the X-ray photon flux ($10^{12}$photons/cm$^2$), the photo-emitted electrons and the lifetime of photo-excited electrons in Au, the relative change in electron density at the Fermi level upon X-ray irradiation is of the order $10^{-8}$ to $10^{-9}$, too small to be detected here. Nevertheless, the achieved resolution ($10^{-5}$) can be enough to observe changes in other materials with reduced density of electrons at the Fermi level as shown below.

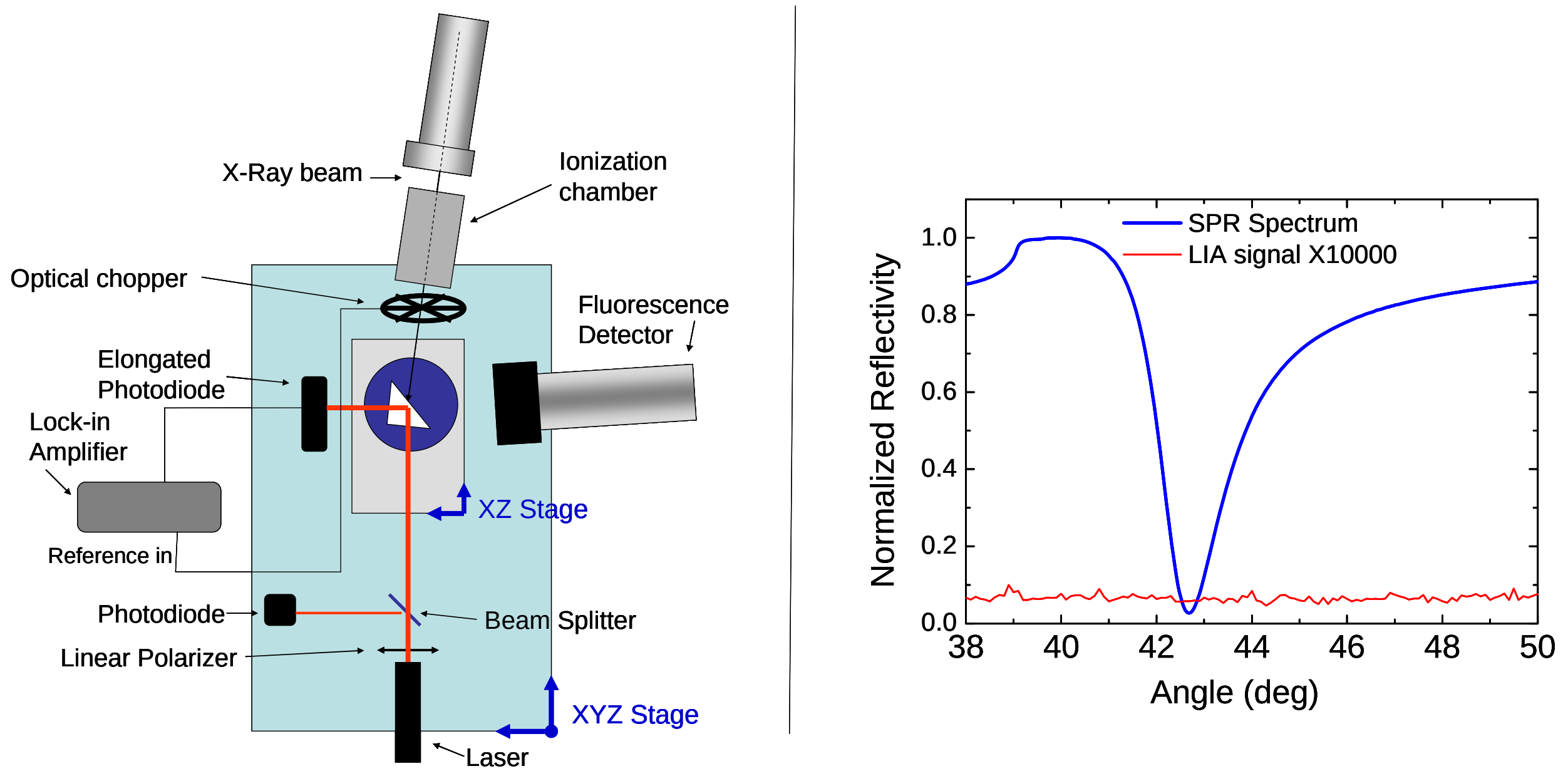

Figure 6. (Color online) (left) Configuration used to detect instantaneous changes induced by the X-rays on the SPR curve; (right) a SPR curve for 45 nm Au film on sodalime glass and the signal detected by the lock-in amplifier in this configuration (multiplied x10000).

The system permits also recording XAS spectra while exciting SPR. Figure 7a shows X-ray absorption near edge structure (XANES) spectra at the Au L3 edge for a 50 nm Au film with and without exciting SPR. In this case, no difference is found up to a resolution of $10^{-3}$ which is the noise level for the integration time we used (30 seconds per point). It is also possible to carry out combined experiments measuring simultaneously X-ray absorption and SPR signal while scanning time, X-rays energy or incidence angle. For instance, we can measure XAS spectra upon excitation of SPR and collect simultaneously the photodiode signal as a function of X-rays energy. To illustrate this procedure (figure 7b), we have recorded simultaneously the XAS signal and the SPR reflectivity at the resonance condition for a bilayer (5nm Fe Oxide / 50 nm Au /silica glass) while scanning the X-ray energy across the Fe K edge (7.112 keV) [21]. There is a slight decrease in the SPR signal when the energy crosses the Fe K-edge. This feature is related to the X-ray absorption of the Fe oxide film although the mechanism is sill not clear: it could be due to a change in the electron population or could be associated with the increase of temperature.

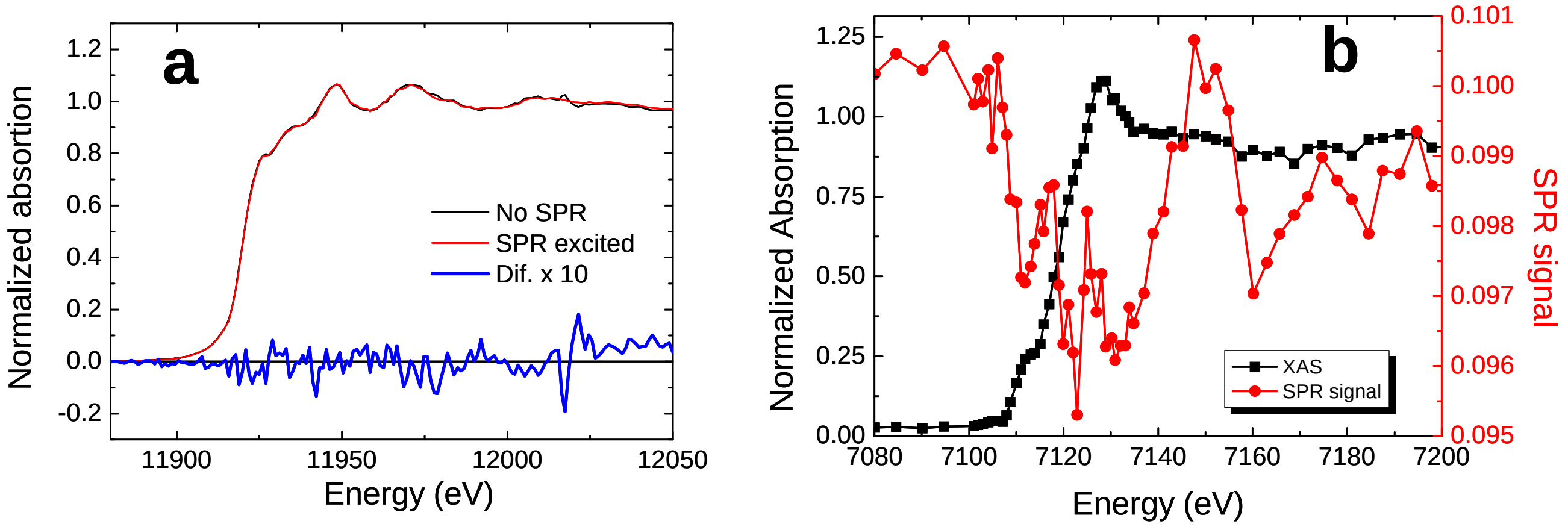

Figure 7. (Color online) (a) XAS spectra measured at the Au L3 edge on a 50 nm Au film with and without excitation of SPR. (b) (black squares) XAS spectra measured a t the Fe Kedge on a 5nm Fe Oxide / 50 nm Au film on a glass substrate upon excitation of SPR and (red circles) the variation in the reflectivity of the SPR curve at the resonance during the energy scan.

In summary, we have developed a set-up combining surface plasmon resonance and X-ray absorption spectroscopy. With this set-up, SPR spectroscopy can be used to explore the interaction of X-rays with matter. Similarly, XAS can be used to study modifications of metallic, semiconductor or dielectric films while SPR is excited. The versatile system allows measuring simultaneously SPR excitation and X-ray absorption while scanning different parameters. The resolution of the measurements is of the order of $10^{-3}$ to $10^{-5}$ depending on the particular experiment. This set-up is available for experiments at the ESRF beamline BM25A – SpLine [22].

## Acknowledgments

The authors acknowledge F. Galvez for assistance during the experiments and J. de la Venta for a critical reading of the manuscript. This work has been supported by the Spanish Spanish Ministerio de Economia y Competitividad (MEC) through the project FIS-2008-06249and MAT2009-14578-C03-02 and Comunidad de Madrid, project NANOBIOMAGNET (S2009/MAT-1726). A. Serrano thanks the CSIC for JAE-Predoctoral fellowship. C. Monton acknowledges AFOSR project number FA 9550-10-1-0409. We acknowledge the European Synchrotron Radiation Facility for provision of synchrotron radiation facilities and the MEC and Consejo Superior de Investigaciones Científicas for financial support (PE-2010 6 0E 013) and for provision of synchrotron radiation facilities and we would like to thank the BM25-SpLine staff for the technical support.